\documentclass[aps,prb,twocolumn,groupedaddress,floatfix,showpacs]{revtex4}
\usepackage{theorem,amssymb,amsbsy,latexsym}
\usepackage{graphicx}% Include figure files

\newcommand{\nd}{{\phantom\dag}}
\newcommand{\mbf}[1]{{\boldsymbol {#1} }}
\newcommand{\cF}{{\cal F}}
\newcommand{\vm}{{\bf m}}
\newcommand{\vs}{{\bf s}}
\newcommand{\vk}{{\bf k}}
\newcommand{\vQ}{{\bf Q}}
\newcommand{\ve}{{\bf e}}
\newcommand{\vsig}{{\mbf \sigma}}
\newcommand{\Ref}[1]{(\ref{#1})}
\newcommand{\eps}{\epsilon}

\begin{document}

%\preprint{}


\title{Phase diagrams of the 2D $t-t'-U$ Hubbard model from an
extended mean field method}


\author{Edwin Langmann}
\email[]{langmann@theophys.kth.se}
%\thanks{}
\affiliation{Mathematical Physics, Physics Department, KTH, AlbaNova, SE-106 91 Stockholm, Sweden} 
\author{Mats Wallin}
\email[]{wallin@theophys.kth.se}
%\thanks{}
\affiliation{Condensed Matter Theory, Physics Department, KTH,
AlbaNova, SE-106 91 Stockholm, Sweden}

\date{June 24, 2004}

\begin{abstract}
It is well-known from unrestricted Hartree-Fock computations that the
2D Hubbard model does not have homogeneous mean field states in
significant regions of parameter space away from half filling.  This
is incompatible with standard mean field theory.  We present a simple
extension of the mean field method that avoids this problem.  As in
standard mean field theory, we restrict Hartree-Fock theory to simple
translation invariant states describing antiferromagnetism (AF),
ferromagnetism (F) and paramagnetism (P), but we use an improved
method to implement the doping constraint allowing us to detect when a
phase separated state is energetically preferred, e.g.\ AF and F
coexisting at the same time.  We find that such mixed phases occur in
significant parts of the phase diagrams, making them much richer than
the ones from standard mean field theory.  Our results for the 2D
$t-t'-U$ Hubbard model demonstrate the importance of band structure
effects.
\end{abstract}

\pacs{71.10.Fd,05.70.Fh,75.50.Ee}
\keywords{Hubbard model, mean field theory, phase diagrams}

\maketitle


\section{Introduction and main results} 

Hubbard-type models in two dimensions have been frequently studied in
the context of high temperature superconductivity and other strongly
correlated systems.\cite{IFT} Despite considerable efforts (for review
see e.g.\ Ref.\ \onlinecite{Dagotto}) there is still need for simple
methods that can contribute to the understanding of the complex
behavior of such models.  In this paper we study an extension of mean
field (MF) theory which allows for the possibility of phase separated
states, in addition to the usual MF states.  We calculate full phase
diagrams for the 2D $t-t'-U$ Hubbard model which, to our knowledge,
are not available in the literature by other methods.

MF theory offers several advantages compared to more complicated
methods like unrestricted Hartree-Fock (HF) theory: It is easy to
implement, not restricted to small system sizes, and can produce phase
diagrams for Hubbard-type models with a limited computational effort.
The disadvantage of {\em standard} MF theory is that it always
predicts translation invariant states everywhere in the phase diagram,
without giving any information about the stability with respect to
fluctuations, or about the stability with respect to competing
non-uniform states.  In 2D Hubbard-type models these problems have
severely restricted the usefulness of the MF
approach,\cite{Hirsch2,LinHirsch} and the MF method is therefore not
widely used.  More correct methods indeed demonstrate that the
qualitative features of the standard MF predictions are restricted to
parts of the phase diagram, e.g., the antiferromagnetic (AF) phase at
half filling.  This suggests that the MF approach is unsatisfactory
and motivates using more complicated methods. However, the more
accurate theoretical methods tend to be computationally demanding and
therefore restricted to very small system sizes.

In this paper we adopt and clarify the extended MF method in Refs.\
\onlinecite{LW1,LW2} and use it to calculate phase diagrams of the 2D
$t-t'-U$ Hubbard model. This method is designed to overcome the
limitation of only producing uniform MF solutions, without increasing
the computational effort. We use the standard mean field
equations,\cite{Penn,Hirsch2,LinHirsch} but we extend them by a method
allowing us to detect possible instabilities towards phase
separation.\cite{LW2} The phase diagrams we thus obtain are much
richer than the ones obtained with conventional MF theory and no
longer in contradiction with unrestricted HF results. In particular,
conventional MF theory for the 2D Hubbard model ($t'=0$) predicts an
AF phase in a finite doping regime around half filling (see Fig.\ 3 in
Ref.\ \onlinecite{Hirsch2}) which is known to be qualitatively wrong.
The phase diagram from the extended MF theory is shown in Fig.~1. It
shows that the AF phase exists only strictly at half filling, and at
finite doping close to half filling no simple translation invariant
state is thermodynamically stable, in agreement with unrestricted HF
theory.\cite{Su,PR,ZG,unrestrHF} While our method does not account for
fluctuations or details of states which are not translational
invariant, it allows to detect frustration in the sense of
incompatibility between MF states and the doping constraint.  Such
frustration suggests interesting physical behavior to be explored by
more sophisticated methods. Our theory should be useful also for other
cases where no other methods are available.

\begin{figure}[ht]
\label{fig1}
\resizebox{!}{6cm}{\includegraphics{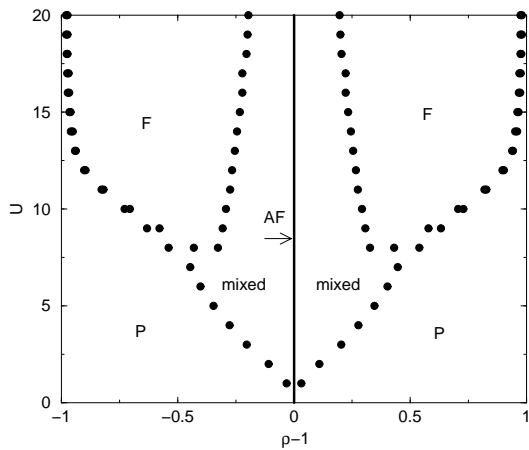}}
\caption{Phase diagram of the 2D Hubbard model as a function of $U$ and
doping $\rho$ for parameters $t=1$ and $t'=0$.  We use Hartree-Fock
theory restricted to ferromagnetic (F), antiferromagnetic (AF) and
paramagnetic (P) states, and we find large mixed regimes where neither
of these translational invariant states is thermodynamically stable.
The results are for $L=60$ and $\beta=1000$ which is practically
indistinguishable from the thermodynamic limit.}
\end{figure}

Our main results are the full phase diagrams for 2D $t-t'-U$ Hubbard
model for $t'=0$ and $t'=-0.35 t$ in Figs.\ 1 and 2, respectively.
They were obtained for a system size so large that they are
practically identical with the thermodynamic limit.  The phase
diagrams are remarkably rich and very different from the corresponding
results from standard MF theory: compare our Fig.~1 with Fig.~3 in
Ref.\ \onlinecite{Hirsch2} and our Fig.~2 with Fig.~1 in Ref.\
\onlinecite{LinHirsch}. Our results demonstrate that mixed phases are
a typical feature of 2D Hubbard-type models: as one changes doping one
never goes directly from one MF phase to another, but there seems
always a finite doping regime with a mixed phase in between. It is
also interesting to note that the qualitative features of the phase
diagram are very sensitive to changes in the next-nearest-neighbor
(NNN) hopping constant $t'$, in qualitative agreement with the
unrestricted HF results.\cite{VVG} In particular, while a pure AF
phase is possible only at half filling for $t'=0$, the AF phase can be
doped by electrons, but not holes, for $t'<0$ at larger values of $U$,
in agreement with previous results obtained with a more complicated
method.\cite{SG} 

\begin{figure}[ht]
\label{fig2}
\resizebox{!}{6cm}{\includegraphics{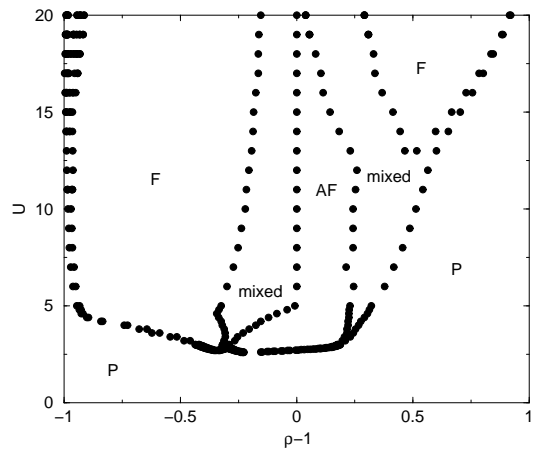}}
\caption{Phase diagram of the 2D Hubbard model as a function of $U$ and
doping $\rho$ for parameters $t=1$, $t'=-0.35$, $L=60$ and
$\beta=1000$, computed as Fig.\ 1.  For large $U$ and $\rho$ close to
zero it becomes numerically difficult to distinguish between the F and
P phase, which is the reason for the fuzzy phase boundaries in this
region of the phase diagram.}
\label{phases}
\end{figure}

The plan of the rest of this paper is as follows. In the next section
we explain and justify our method using physical
arguments. Mathematical details can be found in Sec.~III. Section~IV
contains our conclusion and a summary.

\section{The method}

We now explain our method, concentrating on the point where we deviate
from standard MF theory. Precise mathematical formulas implementing
this method will be given in the next section. As a representative
example we discuss the computation of the phases by our method for the
2D Hubbard model with $U=6$, and $t=1$ and $t'=-0.16$ (see Eq.\
(\ref{hubbard}) below for the precise definitions). One reason for
this choice is that it shows nicely several qualitative features which
can occur in the phase diagram, another that these parameter values
are of interest for high-$T_c$ compounds.\cite{HFSF}

MF theory for the Hubbard model is obtained by restricting HF theory
to translational invariant states describing antiferromagnetism (AF),
ferromagnetism (F) and paramagnetism (P).\cite{Penn,Hirsch2} It would
be straightforward to generalize this and also allow for
charge-density waves, ferrimagnetism etc. One thus starts with three
variational states which all are Slater determinants \cite{Slater}
built of one-particle wave functions which are eigenstates of a mean
field Hamiltonian where the Hubbard interaction is replaced by
external field terms,
\begin{equation} 
|{\rm Slater}\rangle = |X\rangle,\quad X=\mbox{AF, F or P}
\label{Slater} .
\end{equation}
These fields include the the fermions density $\rho$ and the
magnetization which is staggered for AF, constant for F, and zero for
P, and they are determined by the usual Hartree-Fock equations. It is
important to note that the fermion density is fixed in the standard
Slater states, but we use a generalization of Slater's variational
principle to Gibbs states allowing for finite temperature and where the
fermion density is varied by changing a chemical potential $\mu$
(grand canonical ensemble).\cite{BLS,BP,LW2}  We now compute the
Hartree-Fock ground state free energy per site, $\cF_X$, for each of
these states $X=$AF, F and P, as a function of $\mu$.

\begin{figure}[ht]
\label{fig3}
\begin{center}
\includegraphics[width=\linewidth]{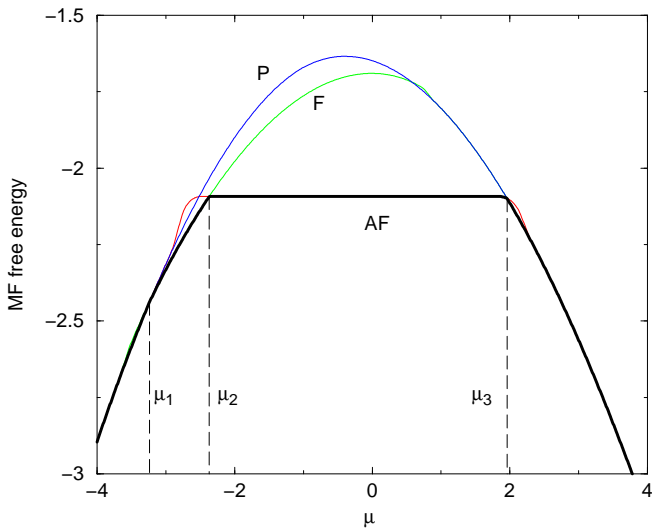}
\end{center}
\caption{Mean field free energy $\cF_X$ of the 2D Hubbard model with
$t=1$, $t'=-0.16$, $U=6$, $L=60$ and $\beta=1000$ as a function of the
chemical potential $\mu$. Shown are the curves for $X=$AF, F and P
(thin lines) and the absolute minimum $\cF_{\rm min}$ (thick
line). The dashed lines indicate the particular values $\mu_{i}$,
$i=1,2,3$, of $\mu$ where the phases change. At these values the
derivative of $\cF_{\rm min}$ has discontinuities, and this leads to
doping regimes with mixed phases; see Fig.\ 4.}
\end{figure}

\begin{figure}[ht]
\label{fig4}
\begin{center}
\includegraphics[width=\linewidth]{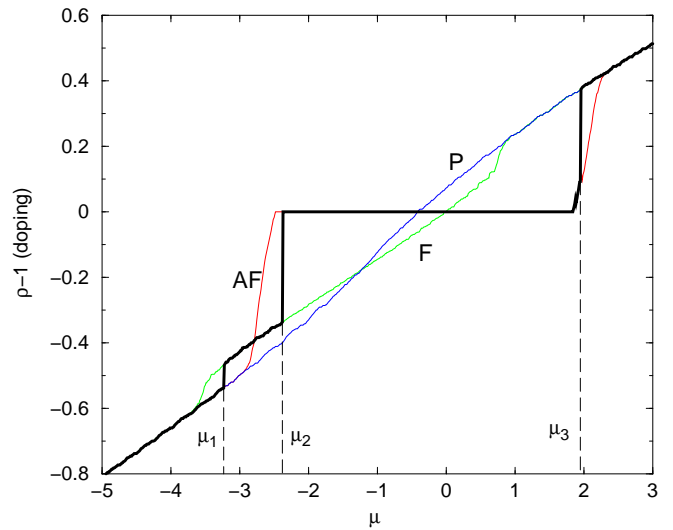}
\end{center}
\caption{Doping $\rho-1$ of the 2D Hubbard model as a function of the
chemical potential $\mu$. The parameters are as in Fig~3 ($t=1$,
$t'=-0.16$, $U=6$, $L=60$ and $\beta=1000$).  The curves are the
derivatives of the corresponding ones in Fig.\ 3. The thick line
determines the mean-field phase diagram, with the discontinuities at
$\mu=\mu_{i}$, $i=1,2,3$ determining doping regions where no pure
phase F, AF or P is thermodynamically stable. The wiggles of the
curves are due to finite size effects which, however, have no effect
on the phase boundaries (this is demonstrated in the inset of Fig.\
5).}
\end{figure}

Figure 3 gives the result for our example.  At fixed value of $\mu$,
the mean field ground state is determined by the minimum,
\begin{equation} 
\cF_{\rm min}=\min_{X={\rm AF,F,P} } \cF_X .
\end{equation}
It is now important to note that the fermion
density can be computed as derivative of the free energy as follows,
\begin{equation} 
\rho-1 = -\frac{\partial \cF_{\rm min}}{\partial \mu} ;
\label{dope} 
\end{equation}
we use conventions such that particle-hole symmetry is manifest, $\rho
-1 \to 1-\rho$ corresponds to $\mu\to -\mu$ and $t'\to -t'$.  From
Figs.~3 and 4 is is obvious that this function $\rho-1$ is, in
general, only piecewise continuous, and it has jumps at the particular
values of $\mu$ where the minimum free energy curve changes, for
example, from the AF to the F curve at the value $\mu=\mu_{2}$. The
physical interpretation of this is as follows. We start at $\mu=0$
where we obviously have the AF ground state and half-filling,
$\rho-1=0$. As we decrease $\mu$, $\rho-1$ remains zero since
$\cF_{\rm AF}$ does not change. This is due to the AF gap: as long as
$\mu$ remains in the gap the fermion density cannot change.  
For large enough $\mu$ values the AF band edge is reached and the
slope of $\cF_{\rm AF}$ starts to decrease. 
However, before this can happen the F free energy has become lower and
taken over: as one decreases $\mu$ the F free energy decreases, and at
a value $\mu=\mu_{2}$ the two curves cross, $\cF_{\rm AF} = \cF_{\rm
F}$ at $\mu=\mu_{2}$. At this point we go from the AF to the F phase.
Since the fermion densities $\rho_X(\mu_{2})-1= -\partial
\cF_X/\partial \mu|_{\mu=\mu_{2}}$ for the states $X=$AF and $X=$F are
different, it is impossible to get a density value in between with
either state.  There is, however, a possibility to realize such a
fermion density with the following state {\em exactly at
$\mu=\mu_{2}$},
\begin{equation} 
|{\rm mixed}\rangle = w|{\rm AF}\rangle + (1-w) |{\rm F}\rangle ,
\label{mixed} 
\end{equation}
with the relative weight $w$ determined by the density as follows,
\begin{equation} 
\rho=w\rho_{\rm AF}(\mu_{2}) + (1-w)\rho_{\rm F}(\mu_{2}),\quad 0 <w <1 . 
\end{equation}

We now discuss the interpretation of this mixed solution. One
possibility is that the system has phase separated and split up into
AF and F regions.\cite{V} Of course, the spatial structure of the
actual state is not available in the MF description by the mixed
state, but it can in principle be calculated using unrestricted
HF. However, since the bulk free energy dominates over the interfacial
free energies in the thermodynamic limit, the mixed state gives an
accurate description of the thermodynamics.  We stress that the
appearance of such a mixed state does {\em not} necessarily mean phase
separation. The effect of the phase boundaries and other possible
states have been excluded in our approximation.  To know the actual
state in the mixed regions thus is beyond our calculation and can be
decided only by doing more work, e.g., using unrestricted HF taking
into account more complicated states. Nevertheless, the {\em
occurrence of such a mixed states proves that no simple translational
invariant state of the kind assumed in our MF ansatz is
thermodynamically stable}.  The mixed regions of the phase diagram are
of particular interest since there the free energy is degenerate and
thus the details of the solution can be strongly affected by
fluctuations, phase boundaries, or details neglected in the model.

It is important to note that there are two further jumps of $\rho$ and
two further corresponding mixed phases: one at $\mu=\mu_{1}$ with F
coexisting with P, and another at $\mu=\mu_{3}$ with AF and P
coexisting. It is also interesting to note that, while for $t'=0$ the
mean field free energies are invariant under the electron-hole
transformation $\mu\to -\mu$, the finite value of $t'=-0.16$ here
leads to a qualitative difference between hole doping ($\mu<0$) and
electron doping ($\mu>0$). As seen in Fig.\ 3, the F state can compete
with the AF state only for $\mu<0$, and this implies that it is
possible to dope the AF state by electrons but not by holes.

We thus see that, even though we restricted Hartree-Fock theory to
simple translation invariant states as in Eq.\ \Ref{Slater}, our way
of treating the doping constraint has implicitly also included the
possibility of having a mixed state as in Eq.\ \Ref{mixed} as
groundstate, and we find that such a mixed state indeed occurs in a
significant part of the doping regime.

We stress that our method to determine the phase boundary does not
increase the computational effort of mean field theory, and it is easy
to do the computations also for large system sizes. Most of our
computations were done for a $L\times L$ lattice with $L=60$.  While
at this values of $L$ some finite size effects are still visible in
the relation between doping $\rho$ and the chemical potential $\mu$
(see Fig.\ 4), the inset in Fig.\ 5 demonstrates that resulting phase
boundaries are practically identical with the ones in the
thermodynamic limit. We also checked that the value $\beta=1000$ we
used for the inverse temperature practically gives the zero
temperature phase boundaries.

\begin{figure}[htb]
\resizebox{!}{6cm}{\includegraphics{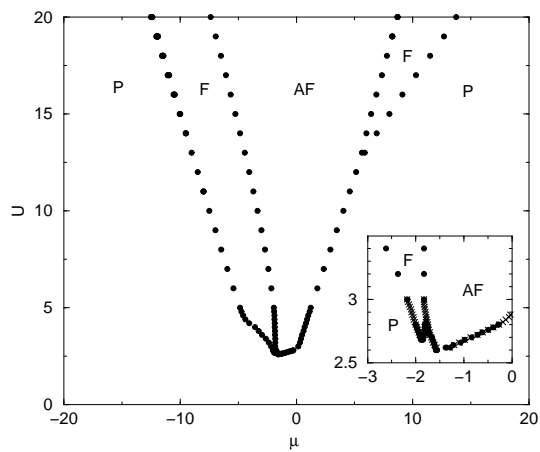}}
\caption{Phases of the 2D Hubbard model as a function of the chemical
potential for the same parameters as in Fig.\ 2 ($t=1, t'=-0.35,
\beta=1000, L=60$).  Inset: Blowup of the region around the minimum of
the phase lines in the main figure, showing interesting fine structure
in the phase diagram.  Also shown is the result from a calculation for
system size $L=120$ (crosses).  The coincidence between results for
two different system sizes demonstrate that $L=60$ is practically
already in the thermodynamic limit.}
\end{figure}

\section{Formalism}

We now give the formal implementation of our method.  We start by
fixing our notation. We consider the 2D Hubbard model defined by the
Hamiltonian
\begin{eqnarray} 
H =  -t \sum_{\langle i,j\rangle  ,\alpha} c^\dag_{i,\alpha} c^\nd_{j,\alpha} 
 - t' \sum_{\langle\!\langle i,j\rangle\!\rangle
 ,\alpha}  c^\dag_{i,\alpha}  c^\nd_{j,\alpha} + {\rm H.c.} \nonumber  \\ -
 \mu \sum_{i,\alpha} c^\dag_{i,\alpha}c^\nd_{i,\alpha}  + U
 \sum_i (n_{i,\uparrow}-\frac12) (n_{i,\downarrow}-\frac12)
\label{hubbard} 
\end{eqnarray}
with the on-site repulsion $U>0$ and the hopping amplitudes $t>0$ and
$t'$ between the nearest neighbor sites $\langle i,j\rangle$ and
next-nearest neighbor (NNN) sites $\langle\!\langle i,j\rangle\!\rangle$
on a square lattice with $L^2$ sites, respectively; the fermion
operators $c^{(\dag)}_{i,\alpha}$ are parameterized by the spin
variable $\alpha=\uparrow,\downarrow$ and lattice sites $i=(i_x,i_y)$
where $i_{x,y} = 1,2,\ldots, L$, and $n_{i,\alpha} =
c^\dag_{i,\alpha}c^\nd_{i,\alpha}$ are number operators, as usual. The
fermion density is
\begin{equation}
\rho = \frac1{L^2} \sum_{i,\alpha} \langle n_{i,\alpha}\rangle
\end{equation}
with $\langle\cdot\rangle$ the ground state expectation value to be
specified below.

We recall that unrestricted Hartree-Fock (HF) theory is formally
obtained by introducing
\begin{equation}
q_i = \langle n_i \rangle,\quad \vm_i = \langle \vs_i \rangle
\label{HF}
\end{equation}
and replacing the Hubbard interaction by external field terms as
follows,
\begin{eqnarray}
n_{i,\uparrow}n_{i,\downarrow} \to \frac14(\vm_i^2 - \rho_i^2)
\nonumber + \frac12(q_i n_i - \vm_i\cdot \vs_i),
\end{eqnarray}
where $\frac12 U \vm_i$ and $\frac12 Uq_i$ are mean fields coupling to
the fermion spin $\vs_i = \sum_{\alpha,\alpha'} c_{i,\alpha}^\dag
\vsig^\nd_{\alpha\alpha'} c^\nd_{i,\alpha'} $ and (local) density $
n_i = \sum_\alpha c_{i,\alpha}^\dag c_{i,\alpha}^\nd, $ respectively;
$\vsig=(\sigma_1,\sigma_2,\sigma_3)$ are the usual Pauli spin
matrices.  This replacement leads to a Hamiltonian describing
non-interacting fermions in external fields, $H\to H_{HF}$ with
\begin{equation}
H_{HF} = \sum_{i}\frac{U}4(\vm_i^2-q_i^2) + \sum_{i,j,\alpha,\alpha'} c^\dag_{i,\alpha}
h^\nd_{i,\alpha,j,\alpha'} c^\nd_{j,\alpha'} \label{HF1}
\end{equation}
where 
\begin{eqnarray} 
h_{i,\alpha,j,\alpha'} = -t_{ij}\delta_{\alpha\alpha'} + \delta_{ij} \Bigl( \frac12 U[ \vm_i\cdot
\vsig_{\alpha\alpha'} + \nonumber \\ 
(q_i-1)\delta_{\alpha\alpha'} ]  -\mu \delta_{\alpha\alpha'} \Bigr) \label{h}
\end{eqnarray}
is a self-adjoint $2L^2\times 2L^2$-matrix which can be interpreted as a
one-particle Hamiltonian.  One now interprets $\langle\cdot\rangle$ in
Eq.\ \Ref{HF} as the expectation value in the ground state of $H_{HF}$ in
Eqs.\ (\ref{HF1},\ref{h}). This yields the HF equations allowing to
self-consistently compute $q_i$ and $\vm_i$ (see e.g.\ Sec.~II in
Ref.\ \onlinecite{unrestrHF}).

We now observe that these HF equations can also be obtained as saddle
point equations $\partial \cF/\partial \vm_i = \partial \cF/\partial
\rho_i=0$ from the free energy function 
\begin{equation}
\cF=-\frac1{\beta L^2} \log {\cal Z}
\end{equation}
where 
\begin{equation} 
{\cal Z} = {\rm Tr} \bigl( {\rm e}^{-\beta H_{HF} }\bigr)
\end{equation}
is the partition function defined by a trace over the fermion Hilbert
space, and $\beta$ is the inverse temperature. A straightforward
computation yields
\begin{equation} 
\cF = \frac{U}{4L^2} \sum_i (\vm_i^2 - \rho_i^2) -
\frac1{L^2}\sum_{\ell=1}^{2L^2} \log\cosh\frac{\beta E_\ell}2 ,\label{F1}
\end{equation}
with $E_\ell$ the eigenvalues of the one-particle Hamiltonian
$h=(h_{i,\alpha,j,\alpha'})$ in Eq.\ \Ref{h}.\cite{LW2}

The physical solution of the HF equations are such that
\begin{equation} 
\cF_{\rm min} \equiv \min_{\vm_i}\, \max_{q_i} \, \cF(\vm_i,q_i); 
\label{F2}
\end{equation}
see Ref.\ \onlinecite{BP} for a mathematical proof or Ref.\
\onlinecite{LW2} for a derivation using functional integrals.  The
corresponding fermion density is then given by Eq.\ (\ref{dope}). We
stress that Eq.\ \Ref{F2}, while {\em implying} standard HF theory, is
not equivalent to it: the standard HF equations can have several
solutions, but Eq.\ \Ref{F2} provides a simple method to solve HF
equations so as to avoid the unphysical solutions: first maximize
$\cF$ with respect to the $q_i$, and then minimize with respect to the
$\vm_i$. In case we restrict HF theory by making a simplifying ansatz
for the mean fields $q_i$ and $\vm_i$ as below, it can happen that one
finds several HF solutions at a fixed value of $\mu$. In this
case one must take the solution minimizing $\cF$.

Mean field theory is obtained from HF by restricting to mean fields
which are invariant under translations by two sites. For the different
states discussed in this paper one further simplifies to
\begin{eqnarray}
{\rm AF:}&& \quad q_i = q,\quad \vm_i = m_{\rm AF} (-1)^{i_x+i_y}
\ve_z \nonumber \\ {\rm F:} && \quad q_i = q,\quad \vm_i = m_{\rm F}
\ve_z \nonumber \\ {\rm P:} && \quad q_i = q,\quad \vm_i = {\mbf 0}
\label{X}
\end{eqnarray}
where $\ve_z$ is the unit vector in $z$-direction.  With this
restrictions it is easy to compute the eigenvalues $E_\ell$ by Fourier
transform. One obtains
\begin{eqnarray} 
{\rm AF:}&& \quad E_{\vk,\pm} = \frac12 \bigl[\eps(\vk) +
\eps(\vk+\vQ)\bigr] + U (q-1) - \mu \pm \nonumber \\ && \qquad \qquad
\quad \pm \sqrt{ \bigl[\eps(\vk) - \eps(\vk+\vQ)\bigr] + (U m_{\rm
AF})^2} \nonumber\\ {\rm F:}&& E_{\vk,\pm} = \eps(\vk)+ U (q - 1 \pm
m_{\rm F})-\mu \nonumber \\ {\rm P:}&& E_{\vk,\pm} = \eps(\vk)+ U (q -
1)-\mu \quad
\end{eqnarray}
where the quantum numbers labeling the eigenvalues are $\ell\equiv
(\vk,\eps)$ with $\eps=\pm$ a band index and $\vk=(k_x,k_y)$ with
$k_{x,y} = (2\pi/L)\times$integer momenta restricted to the Brillouin
zone $-\pi \leq k_{x,y}\leq \pi$; $\vQ=(\pi,\pi)$ is the AF vector,
and
\begin{equation} 
\eps(\vk) = -2t[\cos(k_x)+\cos(k_y)] -4t' \cos(k_x)\cos(k_y)
\end{equation}
is the usual tight binding band relation. Thus the mean field free
energy becomes
\begin{equation} 
\cF_{X} = \frac U4 \bigl( m_X^2 - q^2\bigr) - \frac{1}{L^2}
\sum_{\vk,\varepsilon=\pm}\cosh\frac{\beta}2 E_{\vk,\varepsilon}
\label{FX}
\end{equation}
for $X=$AF, F and P ($m_{\rm P}=0$), where the $\vk$-sum becomes an
integral in the thermodynamic limit $L\to\infty$. The standard mean
field equations (see e.g.\ Sec.~II in Ref.\ \onlinecite{LinHirsch})
are obtained from this from differentiation, $\partial\cF_X/\partial
q=\partial\cF_X/\partial m_{X}=0$. Note that $q=\rho_X$ (fermion
density at fixed $\mu$ in the $X$-state) but, as explained in
Sec.~II, the relation of $\rho_X$ to the system density $\rho$ is
somewhat subtle.

\section{Conclusions and discussion}

In conclusion, we have presented a simple generalization of standard
mean field theory, including the possibility of phase separated mean
field states.  We have presented results for the phase diagram of the
2D $t-t'-U$ Hubbard model, including values of parameters suggested by
the high-$T_c$ materials.  We find that the NNN hopping $t'$
significantly alters the solution.  The resulting rich and nontrivial
phase diagrams show significant qualitative differences between
electron and hole doping.  Moreover, a finite $t'$ suppresses order in
the weak coupling regime, but can have the opposite effect at strong
coupling; see Figs.\ 2 and 5.  Thus the results presented here are
much richer than those obtained by standard MF
theory.\cite{Hirsch2,LinHirsch} The correctness of our method is
justified by mathematical rigorous results.\cite{BP}

We stress that the method presented here does not necessarily produce
accurate solutions to the problem, as is often the case with mean
field theory.  Nevertheless the method provides a useful starting
point for estimating the structure of the phase diagram, providing
cheap guidance for more accurate but costly calculation methods
towards interesting regimes in the phase diagram. 

The simple theory presented here can be straightforwardly generalized
to a number of interesting cases, including more general mean field
states like ferrimagnetism or stripes, and to more complicated models
with additional interaction terms or more bands, etc.

% Surround figure environment with turnpage environment for landscape
% figure
% \begin{figure}
% \end{figure}

% tables should appear as floats within the text
% The ruledtabular enviroment adds doubled rules to table and sets a
% reasonable default table settings.
% Use the table* environment to get a full-width table in two-column
% Add \usepackage{longtable} and the longtable (or longtable*}
% environment for nicely formatted long tables. Or use the the [H]
% placement option to break a long table (with less control than 
% \begin{table}%[H] add [H] placement to break table across pages
% \begin{ruledtabular}
% Lines of table here ending with \\
% \end{tabular}
% \end{ruledtabular}
% \end{table}

% Surround table environment with turnpage environment for landscape
% table
% \begin{table}
% \begin{ruledtabular}
% \end{tabular}
% \end{ruledtabular}
% \end{table}

%\appendix
%\section{}

\begin{acknowledgments}
We thank Manfred Salmhofer for helpful discussions.  This work was
supported by the Swedish Science Research Council~(VR) and the G\"oran
Gustafsson Foundation.
\end{acknowledgments}

\bibliography{LW_HF}
%%\bibliography{apssamp}% Produces the bibliography via BibTeX.

\end{document}